\documentclass[fleqn,usenatbib]{mnras}
\usepackage{newtxtext,newtxmath}
\usepackage[T1]{fontenc}
\usepackage{graphicx}
\usepackage{amsmath}

\title[Disc-Jet Couplings in XRBs]{
Accretion Disc-Jet Couplings in X-ray Binaries}
\author[P.-X. Shen \& W.-M. Gu]{
Pei-Xin Shen $^{1,2}$
Wei-Min Gu $^{1}$
\thanks{Contact e-mail: 
\href{mailto:guwm@xmu.edu.cn}{guwm@xmu.edu.cn}}\\
$^{1}$Department of Astronomy, 
Xiamen University, Xiamen, Fujian 361005, China\\
$^{2}$Institute for Interdisciplinary Information Sciences,
Tsinghua University, Beijing 100084, China}

\date{Accepted XXX. Received YYY; in original form ZZZ}
\pubyear{2020}

\begin{document}
\label{firstpage}
\pagerange{\pageref{firstpage}--\pageref{lastpage}}
\maketitle

\begin{abstract}
When the matter from a companion star is accreted towards the central
compact accretor, i.e. a black hole (BH) or a neutron star (NS), an
accretion disc and a jet outflow will form, providing bight X-ray and
radio emission, which is known as X-ray binaries (XRBs). In the
low/hard state, there exist disc-jet couplings in XRBs, but it remains
uncertain whether the jet power comes from the disc or the central
accretor. Moreover, BHXRBs have different properties compared with
NSXRBs: quiescent BHXRBs are typically two to three orders of
magnitude less luminous than NSXRBs in X-ray, whereas BHXRBs are more
radio loud than NSXRBs. In observations, an empirical correlation has
been established between radio and X-ray luminosity, $L_{\rm R}
\propto L_{\rm X}^b$, where $b\sim 0.7$ for BHXRBs and $b \sim 1.4$
for non-pulsating NSXRBs. However, there are some outliers of
BHXRBs showing unusually steep correlation as NSXRBs at higher
luminosities. In this work, under the assumption that the origin of
jet power is related to the internal energy of the inner disc, we
apply our magnetized, radiatively efficient thin disc model and the
well-known radiatively inefficient accretion flow model to NSXRBs and
BHXRBs. We find that the observed radio/X-ray correlations in XRBs can
be well understood by the disc-jet couplings.
\end{abstract}


\section{Introduction} \label{Sec:Introduction}

Most stars in the Universe are in the form of binary systems. When one
of these two stars evolves into a black hole or a neutron star, its
companion star may provide materials to be accreted by the compact
object and form an accretion disc and a jet outflow. This process will
release large amounts of gravitational energy and radiate bright X-ray
from the disc, which is known as X-ray binaries (XRBs), together with
radio emission from the jet. Generally, XRBs are classified into black
hole X-ray binaries (BHXRBs) and neutron star X-ray binaries (NSXRBs)
according to the type of central accretor. The radiation properties
are quite different between BHXRBs and NSXRBs: in the
quiescent state (QS), BHXRBs are typically two to three orders of
magnitude less luminous than NSXRBs
\citep{Fender03, Narayan08}, but are more radio loud than NSXRBs,
which indicates that the jet power in BHXRBs will be more powerful
than that in NSXRBs; while in high/soft state (HSS), jet activities
fade away for both BHXRBs and NSXRBs. In summery, we have the
following relation for radio luminosities (related to jet activities)
of different states:
\begin{equation} \label{Eqn:L}
  L_{\rm R}^{\rm BH} ({\rm QS}) > L_{\rm R}^{\rm NS} ({\rm QS})
  > L_{\rm R} ({\rm HSS}) \ .
\end{equation}

In addition, how the relativistic jet forms, and how much power it
carries are the key questions. It has long been suggested that
relativistic jets could be powered not by the accretion flow but by
the spin of the black hole \citep{Blandford77, Narayan12}. However,
some scientists \citep{Blandford82, Fender10} believe that the jet
power mainly comes from the accretion disc rather than the spin of the
black hole due to the disc-jet coupling phenomenon from observation.
\citet{Corbel03} discovered that, over four orders of magnitude in
X-ray luminosity, the relation between radio and X-ray
luminosity in the low/hard state (LHS) of the BHXRB GX 339-4 has the
form of $L_{\mathrm{R}} \propto L_{\mathrm{X}}^b$, where $b\sim0.7$
for $L_{\mathrm{X}}$ in the 3-9 keV range. \citet{Gallo03} analysed a
large sample of BHXRBs (including V404 Cyg, A0620-00) and extended
this correlation down to the quiescent level, which is supposed to be
a universal correlation between the radio and X-ray luminosity for
BHXRBs. However, in recent years, some newly detected BHXRBs sources
(e.g., H1743-322, Swift J1753.5-0127 and MAXI J1659-152) are found to
lie significantly below the standard correlation (hereafter
``outliers''). \citet{Coriat11} found BHXRB H1743-322 with a steeper
correlation index of $b\sim1.4$, but at lower luminosity it may return
to the standard correlation, which presents an evidence for two
different tracks for BHXRBs. Nowadays, the origin of these
outliers are still under debate: on the one hand, \citet{Gallo14,
Gallo18} argued that there is no statistical evidence for black hole
systems following different tracks in the radio/X-ray luminosity
plane; on the other hand, in the meantime, \citet{Islam18} claimed
that H1743-322 still follows different tracks from GX 339-4 when
considering radio/bolometric flux correlations, which suggests that
the accretion disc may undergo a morphological transition from
radiatively inefficient to radiatively efficient flow
\citep{Koljonen19}.

As for NSXRBs, they are generally less radio loud for a given X-ray
luminosity (regardless of mass differences for NS and BH) and they do
not appear to show the strong suppression of radio emission in HSS
that we observe in BHXRBs. However, recent studies
\citep{Tudor17,Tetarenko18} show that different NSXRB classes follow
different correlations and normalizations in the radio/X-ray plane
(possibly owing to the limited range of X-ray luminosities).
Specifically, \citet{Migliari06} found that the correlation is
consistent with $L_{\mathrm{R}}\propto L_{\mathrm{X}}^{1.4}$ including
data taken from Aql X-1 and 4U 1728-34, over only one order of
magnitude in X-ray luminosity. \citet{Tetarenko16} reported the third
individual NSXRB simultaneous radio and X-ray measurements of
EXO1745-248 and obtained its radio/X-ray correlation in the form of
$L_{\mathrm{R}}\propto L_{\mathrm{X}}^{1.68}$. While pulsating
systems such as accreting millisecond X-ray pulsars (AMXPs) and
transitional millisecond X-ray pulsars (tMSPs) tend to show a
shallower correlation $b\sim0.7$ \citep{Deller15}. Especially for the
strongly magnetized accreting X-ray pulsar, they are likely to occupy
a different radio/X-ray plane region with fainter radio luminosity
\citep{vandenEijnden18,vandenEijnden19}. In this paper, we will focus
on the non-pulsating NSXRBs with the steeper correlation $b\sim 1.4$,
since the strong large-scale magnetic fields in X-ray pulsars can
disrupt the inner disc and therefore the accretion disc may not extend
to the NS surface.

It has been commonly believed that the correlation indexes of
$\sim1.4$ and $\sim0.7$ correspond to a radiatively efficient and a
radiatively inefficient accretion flows, respectively \citep{Soleri10,
Coriat11, Jonker12, Ratti12, Russell15, Plotkin17}. Table~\ref{Tab:Disc}
present a concise comparisons among different accretion disc models.
The luminous hot accretion flow (LHAF) proposed by \citet{Yuan01} and
some magnetic accretion disc corona models \citep{Merloni03} can be
applied to the radiatively efficient track. As for the radiatively
inefficient track, it can be explained by the advection-dominated
accretion flow (ADAF) \citep{Narayan94,Narayan95,Abramowicz95}
or some X-ray jet-dominated models \citep{Markoff03,
Markoff05}. In previous researches \citep{Narayan97a, Narayan98,
Narayan12}, the ADAF model has been widely applied to BHXRBs, since
the existence of horizon will not interrupt the process of advection,
and the accreted matter can pass the horizon smoothly at a speed of
light. However, things are quite different for NSXRBs: due to the
substantial surface, there remains a caveat that the matter falling
towards NS may be rebounded back, blocking up the advection-dominated
flow. In this case, the ADAF model may not be well applicable to
NSXRBs. Nevertheless, in order to explain the high frequency spectrum
of NSXRBs, the disc ought to be optically thin. Except for ADAF,
\citet{Shapiro76} established an accretion disc model called
Shapiro-Lightman-Eardley (SLE) disc, which is both optically
and geometrically thin. Unfortunately, the original SLE disc was
identified to be thermally unstable \citep{Pringle76,Piran78}, while
later on \citet{Yu15} found that the SLE disc could be stable when the
small-scale turbulent magnetic pressure is strong enough,
which is called the magnetized-SLE disc model (hereafter the M-thin
disc). They showed that the M-thin disc is more likely to
exist than other radiatively efficient disc models, and briefly
indicated that it could be used to uncover the disc-jet couplings of
XRBs.

In this paper, we take a further step to formulize the
disc-jet coupling correlations based on the M-thin disc and the ADAF
model, by assuming the jet power is related to the internal energy at
the inner radius of the accretion disc.
In Section \ref{Sec: Equations}, we introduce the
equations of the M-thin disc and the ADAF model respectively, along
with deducing the analytical radio/X-ray correlations with the aid of the
aforementioned jet power assumption. The results for the theoretical
temperature properties, as well as the comparison between theoretical
regions with 35 observational XRB sources are presented in Section
\ref{Sec: Results}. Our conclusions and discussion are given in
Section \ref{Sec: Conclusions}.

\begin{table*}
  \caption{\label{Tab:Disc} Properties of different accretion disc models}
  \centering
  \begin{tabular}{cccccccc}
  \hline
  Models & Optically & Geometrically & Thermally & Pressure      & Cooling   & Inner Disc $T_\mathrm{i}$ & $L_{\mathrm{R}} \propto L_{\mathrm{X}}^b$ \\ \hline
  SSD    & thick     & thin          & stable/unstable    & gas/radiation & radiation & $\sim 10^7$ K     & N/A\\
  SLE    & thin      & thin          & unstable  & gas           & radiation & $\sim 10^9$ K     & N/A\\  
  ADAF   & thin      & thick         & stable    & gas           & advection & $\sim 10^{11}$ K  & $\sim0.7$\\ 
  LHAF   & thin      & thick         & stable/unstable  & gas    & radiation & $\sim 10^{10}$ K  & $\sim1.4$\\ 
  M-thin & thin      & thin          & stable    & gas/magnetic  & radiation & $\sim 10^9$ K     & $\sim1.4$\\ \hline
  \end{tabular}
\end{table*}

\section{Equations} \label{Sec: Equations}

\subsection{M-thin disc}

The traditional SLE model follows these assumptions:
\begin{itemize}
  \item The disc is assumed to rotate at $\Omega_\mathrm{K}$, which is
the Keplerian angular velocity. We adopt that the general relativistic
effect of the central accretor is simulated by the well-known
pseudo-Newtonian potential introduced by \citet{Paczynsky80}, i.e.,
$\Phi=-GM/(R-R_\mathrm{s})$, where $M$ is the mass of accretor, $R$ is
the radius, and $R_\mathrm{s}=2GM/c^2$ is the Schwarzschild radius.
Therefore, $\Omega=\Omega_\mathrm{K}=(GM/R)^{1/2}/(R-R_\mathrm{s})$;
  \item The vertical half thickness of the flow $H$ is smaller than
the radius $R$, i.e. $H(R)=c_\mathrm{s}/\Omega_\mathrm{K}\lesssim R$,
where $c_\mathrm{s}= \sqrt{p/\rho}$ is the isothermal local sound
speed, with $p$ and $\rho$ being the pressure and mass density
respectively;
  \item The kinematic viscosity coefficient $\nu$ is expressed as
$\nu=\alpha c_\mathrm{s} H$, where $\alpha$ is the constant viscosity
parameter.
\end{itemize}
Considering a vertically average, steady axis-symmetric accretion flow,
we use unified steady state accretion flow equations introduced by
\citet{Chen95}:
\begin{equation}
  \nu\Sigma=\frac{\dot{M}}{3\pi} f g^{-1} \ ,
\end{equation}
where $\dot{M}$ is accretion rate, surface density $\Sigma=2H\rho$,
$f=1-j/(\Omega R^2)$, $g=-(2/3)(\mathrm{d}\ln\Omega_{\rm K}/\mathrm{d} \ln R)$
and $j$ is the specific angular momentum. When $\Omega=\Omega_\mathrm{K}$,
$g=1 + (2/3)/(R/R_{\rm s} - 1)$, and
\begin{equation}
  \dot{M}=-2\pi R\Sigma v_\mathrm{R} \ ,
\end{equation}
where $v_\mathrm{R}$ is the radial velocity. To construct the energy
conservation equation, we introduce the following assumptions:
\begin{itemize}
  \item The dissipation energy goes into the ions;
  \item The energy is transferred from the ions to the electrons
        through Coulomb coupling;
  \item The electrons are cooled by bremsstrahlung process.
\end{itemize}
Based on these assumptions, the energy equations of the ions and electrons
can be respectively written as
\begin{equation} \label{Eqn:MthinEnergy}
  Q_\mathrm{+}= Q_\mathrm{-};\; Q_\mathrm{-}=\Lambda_{\mathrm{ie}};\;
  \Lambda_{\mathrm{ie}}=Q_{\mathrm{brems}} \ ,
\end{equation}
where $Q_\mathrm{+}$ is the viscous heating rate per unit area,
$Q_\mathrm{-}$ is the cooling rate per unit area, $\Lambda_{\mathrm{ie}}$
is the energy transfer rate from the ions to the electrons per unit area
and $Q_{\mathrm{brems}}$ is the sum of radiatively cooling rate
of electron per unit area. $Q_\mathrm{+}$ can be written as
\begin{equation}
  Q_\mathrm{+}=\frac{3\dot{M}}{4 \pi}\Omega^2 f g \ .
\end{equation}
$\Lambda_{\mathrm{ie}}$ is given by \citet{Kato08}:\begin{equation}
\Lambda_{\mathrm{ie}}=\frac{3}{2}\nu_\mathrm{E} \frac{\Sigma
k_{\rm B}(T_\mathrm{i} - T_\mathrm{e})}{m_\mathrm{p}} \ ,
\end{equation}
where $k_{\rm B}$ is the Boltzmann constant, $m_\mathrm{p}$ is the proton mass,
and $T_\mathrm{i}$ and $T_\mathrm{e}$ are the ion and electron temperatures,
respectively. $\nu_{\mathrm{E}} = 2.4\times 10^{21} \ln \Lambda
~\rho\, T_\mathrm{e}^{-3/2}$, $\ln \Lambda$ is the Coulomb logarithm,
here we take $\ln \Lambda=15$.
Bremsstrahlung loss per unit volume from an electron gas is \citep{Svensson84}:
\begin{equation}
  q_{\mathrm{brems}} = \alpha_\mathrm{f} r_\mathrm{e}^2 m_\mathrm{e} c^3 n
  _\mathrm{e}^2 f_{\mathrm{brems}} \ ,
\end{equation}
where $\alpha_\mathrm{f}=e^2/2\pi\epsilon_0 hc \approx 1/137$
is the fine-structure constant, $r_\mathrm{e}=e^2/m_\mathrm{e}c^2$
is the classical electron radius, and $n_\mathrm{e}$ is the electron number
density. Here we assumed $n_\mathrm{i}=n_\mathrm{e}=\rho/m_\mathrm{p}$.
$f_{\mathrm{brems}}$ is the dimensionless radiation rate due to
relativistic bremsstrahlung, which can be decomposed into a part due
to proton-electron collisions, $f_{\mathrm{ep}}$, and that due to
electron-electron collisions, $f_{\mathrm{ee}}$, as
\begin{equation}
  f_{\mathrm{brems}}=f_{\mathrm{ep}}+f_{\mathrm{ee}} \ ,
\end{equation}
where $f_{\mathrm{ep}}$ and $f_{\mathrm{ee}}$ are approximated according
to the following forms \citep{Svensson84}:
\begin{equation}
  f_{\mathrm{ep}}(\Theta_\mathrm{e})=\left\{\begin{array}{rl}
  12\Theta_\mathrm{e}\left(\mathrm{log}(2\,\eta_\mathrm{E}\,
  \Theta_\mathrm{e}+0.42)+\frac{3}{2}\right), & \Theta_\mathrm{e}> 1\\[2pt]
  \frac{32}{3}\left(\frac{2}{\pi}\right)^{1/2}\Theta_\mathrm{e}^{1/2}
  (1+1.78\Theta_\mathrm{e}^{1.34}), & \Theta_\mathrm{e}<1 \ ,
  \end{array}\right.
\end{equation}
\begin{equation}
  f_{\mathrm{ee}}(\Theta_\mathrm{e})=\left\{\begin{array}{rl}
  24\Theta_\mathrm{e}\left(\mathrm{log}(2\,\eta_\mathrm{E}\,\Theta_\mathrm{e})
  +\frac{5}{4}\right), & \Theta_\mathrm{e}> 1\\[2pt]
  \frac{20}{9\pi^{1/2}}(44-3\pi^2)^{1/2}\Theta_\mathrm{e}^{3/2}\times\\
  (1+1.11\Theta_\mathrm{e}+\Theta_\mathrm{e}^2-1.25\Theta_\mathrm{e}^{2.5}),
  & \Theta_\mathrm{e}<1 \ ,\end{array}\right.
\end{equation}
where $\eta_\mathrm{E} = \mathrm{e}^{-\gamma_\mathrm{E}}$ and
$\gamma_\mathrm{E} \approx 0.5772$ is the Euler constant,
$\Theta_\mathrm{e} = k_{\rm B}\,T_\mathrm{e}/m_\mathrm{e}c^2$ the
dimensionless electron temperature. The cooling rate per unit surface
area of discs by bremsstrahlung is given by
\begin{equation}
  Q_{\mathrm{brems}}=2H q_{\mathrm{brems}} \ ,
\end{equation}
where we neglect Compton amplification. Since the radiation pressure
is negligible in optically thin discs, and we take the magnetic pressure
into account in the M-thin disc, the total pressure $p$ is expressed as
\begin{equation}
  p = p_{\mathrm{gas}} + p_{\mathrm{mag}} \ ,
\end{equation}
where $p_{\mathrm{gas}} = \beta p$ is the gas pressure and
$p_{\mathrm{mag}} =(1-\beta)p$ is the magnetic pressure. In this
paper, we take $\beta=0.5$ in all calculations, which was
previously adopted for global ADAF solutions \citep{Narayan97b},
and was used in \citet{Yu15} for the thermal stability analyses.

The gas is assumed to consist of protons and electrons,
so the gas pressure $p_{\mathrm{gas}}$ is written as
\begin{equation}
  p_{\mathrm{gas}} = \frac{\rho\,k_{\rm B}}{\mu\,m_\mathrm{p}}(T_\mathrm{i}
  + T_\mathrm{e}) \ ,
\end{equation}
where $\mu$ is the mean molecular weight, we set $\mu = 1$ in this paper.

\subsection{ADAF}

Different from the M-thin disc, the ADAF is advection-dominated,
thus we cannot expect
angular velocity to be Keplerian. The original radial momentum equation
in fluid dynamics is
\begin{equation}
  v_\mathrm{R} \frac{\mathrm{d}v_\mathrm{R}}{\mathrm{d}R}-\Omega^2R
  = -\Omega^2_\mathrm{K}R-\frac{1}{\rho}\frac{\mathrm{d}}{\mathrm{d}R}
  (\rho c^2_\mathrm{s}) \ .
\end{equation}
We assume $v_\mathrm{R}\propto R^{-1/2}$, $\rho c_\mathrm{s}^2\propto
R^{-5/2}$, the radial momentum equation is reduced to an algebraic form
\begin{equation}
  -\frac{1}{2}\frac{v_\mathrm{R}^2}{R}-\Omega^2 R
  = -\Omega_\mathrm{K}^2 R+\frac{5}{2}\frac{c_\mathrm{s}^2}{R} \ .
\end{equation}
$v_\mathrm{R}$ is negligible compared with $c_\mathrm{s}$,
then we obtain the angular velocity in ADAFs: 
\begin{equation}
  \Omega^2 = \Omega_\mathrm{K}^2 - \frac{5}{2}\frac{c_\mathrm{s}^2}{R^2} \ .
\end{equation}

Except for the angular velocity, another difference between the ADAF
and the M-thin disc is the energy equation: 
\begin{equation}
  Q_\mathrm{+}= Q_\mathrm{-};\; Q_\mathrm{-}=\Lambda_{\mathrm{ie}}
  +Q_{\mathrm{adv}};\; \Lambda_{\mathrm{ie}}=Q_{\mathrm{brems}} \ .
\end{equation}
The additional term $Q_{\mathrm{adv}}$ is the cooling rate per unit area
by advection, which is given by \citep{Chen93}:
\begin{equation}
  Q_{\mathrm{adv}} = \frac{\dot{M}}{2 \pi R^2} \frac{p}{\rho} \xi \ ,
\end{equation}
where $\xi= -[(4-3\beta)/(\Gamma_3-1)](\mathrm{dln} T/\mathrm{dln} R)
+ (4-3\beta)(\mathrm{dln} \Sigma/\mathrm{dln} R)$,
$\Gamma_3 = 1 + (4-3\beta)(\gamma-1)/[\beta + 12(\gamma-1)(\beta-1)]$,
and $\gamma$ is the ratio of specific heats. For simplicity,
we take $\xi=1$ into calculation.

Since the ADAF is a geometrically thick accretion disc, the outflow in the
inner region ($R<100 R_\mathrm{s}$) cannot be negligible. We assume
the accretion rate varies as
\begin{equation}
  \dot{M} = \dot{M}_{*} \frac{R}{100 R_\mathrm{s}},\quad R<100 R_\mathrm{s} \ ,
\end{equation}
where $\dot{M}_{*}$ is the accretion rate at $R=100 R_\mathrm{s}$.
When $R>100 R_\mathrm{s}$, $\dot{M} = \dot{M}_{*}$ at any given $R$.

Other assumptions and equations for the ADAF remain the same as the M-thin disc.

\subsection{Analytical correlations}

Except for numerical calculation, \citet{Fender03} and
\citet{Migliari06} obtain the indexes of radio/X-ray correlations via
algebraic calculation. Analogously, we can also deduce these slope
indexes from our model. For simplicity, in the following discussion,
the accretion rate is given in the Eddington unit, i.e.
$\dot{m}\equiv\dot{M} / \dot{M}_{\mathrm{Edd}}~(\dot{M}_
{\mathrm{Edd}}\equiv L_{\mathrm{Edd}}/\eta c^2)$, where
$L_{\mathrm{Edd}}$ is the Eddington luminosity of $1.3\times
10^{38}(M/M_\odot) \; \mathrm{erg/s}$, $\eta$ is the accretion
efficiency of $\sim1/16$ in the pseudo-Newtonian potential. The radius
is also denoted in the Schwarzschild unit, i.e. $r=R/R_\mathrm{s}$.

Theoretically, the ADAF is usually applied to the inner region of the
accretion disc for BHXRBs, the maximum ion temperature is around
$10^{11}$K \citep{Narayan08}. Through our calculation, the maximum ion
temperature of M-thin discs is significantly lower than ADAFs, which
is around $10^9$K (see Figure~\ref{Fig:T-R} and Section~\ref{Sec:
Results} for details). Besides, an optically thick but
geometrically thin standard Shakura-Sunyaev disc (SSD) was proposed to
describe XRBs in HSS, whose maximum ion temperature is around
$10^7$K \citep{Shakura73}. To sum up, the relation for temperatures
of different accretion disc models are shown as:
\begin{equation} \label{Eqn:T}
T_{\mathrm{ADAF}}> T_{\mathrm{M-thin}} > T_{\mathrm{SSD}} \ ,
\end{equation}
Comparing Equation~\eqref{Eqn:T} with Equation~\eqref{Eqn:L}, we find
that there remain some connections between jet power $L_{\mathrm{jet}}$
and internal energy rate $\dot{U}$ (since materials are accreting,
energy per unit time is a meaningful term). The internal energy rate
$\dot{U}$ can be estimated by
\begin{equation}
  \dot{U}=\frac{3 k\,T_{\mathrm{i}}^\mathrm{max}}{2\mu\,m_\mathrm{p}}\dot{M}\ .
\end{equation}
If we assume the radio luminosity is related to internal energy rate,
we can obtain the radio luminosity as
\begin{equation}
  L_{\mathrm{R}}=f_* \dot{U} \ ,
\end{equation}
where $f_*$ is the transferred fraction from internal energy rate to
radio luminosity. Note that the value of $f_*$ should be
relatively small since the fraction of the disc internal energy
converting to the jet power is already a small value. Moreover,
the portion of the jet emitting radio is also very small. Through
fitting with the observational data, we set $f_*=10^{-6.5}$ for both
M-thin discs and ADAFs. The X-ray luminosity can be obtained by
integration:
\begin{equation}
  L_{\mathrm{X}}=2 \pi \int Q_{\mathrm{brems}}R\,\mathrm{d}R \ .
\end{equation}
Since the X-ray luminosity is dominated by the inner region, we integrate
it from the inner stable circular orbit $3R_\mathrm{s}$ to 100$R_\mathrm{s}$.

In ADAFs, the accretion flow is radiatively inefficient,
the X-ray luminosity follows \citep{Rees82}
\begin{equation}
  L_{\mathrm{X}}\propto\dot{m}^2 \ .
\end{equation}
The temperature of ADAFs is nearly independent of $\dot{m}$ (see
Figure~\ref{Fig:T-m} and Section~\ref{Sec: Results} for details),
\begin{equation}
  L_{\mathrm{R}}=f_* \dot{U}=f_*\frac{3 k\,T_{\mathrm{i}}
  ^{\mathrm{max}}}{2 \mu\,m_\mathrm{p}} \dot{M}\propto \dot{m} \ .
\end{equation}
Thus the theoretical radio/X-ray correlation is given by
$L_{\mathrm{R}}\propto L_{\mathrm{X}}^{0.5}$.
As for the M-thin disc, the accretion flow is radiatively efficient,
and the X-ray luminosity is proportional to the accretion rate
\begin{equation}
  L_{\mathrm{X}}\propto\dot{m} \ .
\end{equation}
Different from ADAFs, the temperature of M-thin discs is varied with
accretion rate, we should deduce the correlation via basic
energy equations \eqref{Eqn:MthinEnergy}. The viscous heating rate
per unit area $Q_\mathrm{+}$ can be simplified as
\begin{equation}
  Q_\mathrm{+}=\dot{m}F_1(r) \ ,
\end{equation}
and the electron radiatively cooling rate per unit area $Q_{\mathrm{brems}}$
can be estimated by
\begin{equation}
  Q_{\mathrm{brems}}=\Lambda_{\mathrm{ie}}=\dot{m}^2\frac{(T_\mathrm{i}
  -T_\mathrm{e})}{T_\mathrm{e}^{3/2}(T_\mathrm{i}+ T_\mathrm{e})^{5/2}}F_2(r)
  \approx \frac{\dot{m}^2}{T_\mathrm{i}^2}F_2(r) \ ,
\end{equation}
where $F_1(r), F_2(r)$ denote some complicated functions only
varied with $r$. The thermal stability forces the M-thin disc to
maintain $Q_\mathrm{+}=Q_{\mathrm{brems}}$ at any given $r$, which
makes the temperature follow $T_\mathrm{i}\propto \dot{m}^{0.5}$ (this
correlation is close to our numerical result
$T_{\mathrm{i}}^\mathrm{max}\propto \dot{m}^{0.7}$ in
Figure~\ref{Fig:T-m}). Hence, the radio luminosity is given by
\begin{equation}
  L_{\mathrm{R}}=f_* \dot{U}=f_*\frac{3 k\,T_{\mathrm{i}}^\mathrm{max}}
  {2\mu\,m_\mathrm{p}} \dot{M}\propto \dot{m}^{0.5}\times \dot{m}=\dot{m}^{1.5}\ .
\end{equation}
Therefore, the radio/X-ray correlation in the M-thin disc is approximately
in the form
of $L_{\mathrm{R}}\propto L_{\mathrm{X}}^{1.5}$.

\section{Results} \label{Sec: Results}

\subsection{Temperature properties}

\begin{figure}
	\includegraphics[width=\columnwidth]{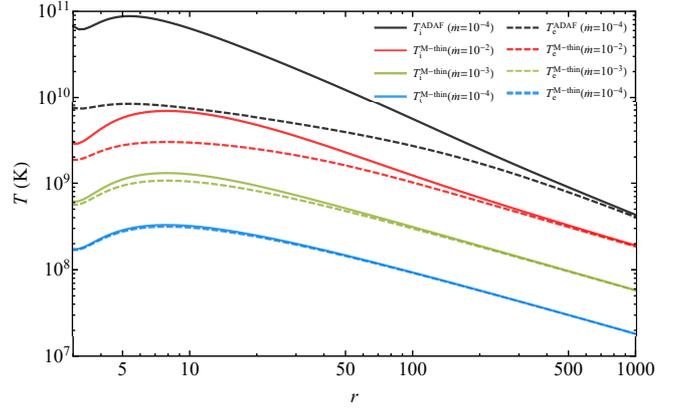}
  \caption{Radial variations of the ion temperature $T_\mathrm{i}$
  (solid) and the electron temperature $T_\mathrm{e}$ (dashed) for
  $\dot{m}$ = 10$^{-4}$ (blue), 10$^{-3}$ (green), and 10$^{-2}$ (red)
  in M-thin discs ($M=1.4M_\odot$), as well as for ADAFs ($M=10M_\odot$,
  black), where $\alpha=0.1$, $j=1.8cR_\mathrm{s}$ in both models.}
  \label{Fig:T-R}
\end{figure}

Based on the equations introduced above, we can calculate the
temperature distributions for M-thin discs and ADAFs, respectively. We
take $M=1.4M_\odot$ for M-thin discs and $M=10M_\odot$ for ADAFs.
Figure~\ref{Fig:T-R} shows the radial variations of the ion and
electron temperatures for M-thin discs and ADAFs with $\alpha=0.1$,
$j=1.8cR_\mathrm{s}$. We can see that the ion temperature
$T_{\mathrm{i}}$ (solid line) is significantly higher than the
electron temperature $T_{\mathrm{e}}$ (dashed line) in the inner
region of the disc. We set $\dot{m}$ = 10$^{-4}$ (blue), 10$^{-3}$
(green), and 10$^{-2}$ (red) for M-thin discs, and $\dot{m}_{*}$ =
10$^{-4}$ (black) for ADAFs (recall that $\dot{m}_{*}$ is the
accretion rate given in Eddington unit at $R=100 R_\mathrm{s}$). It is
seen that the ion temperature of the ADAF is significantly higher than
the M-thin disc, and they both have maximum ion temperature
$T_{\mathrm{i}}^\mathrm{max}$ and maximum electron temperature
$T_{\mathrm{e}}^\mathrm{max}$ in the inner region. When $\alpha=0.1$,
$j=1.8cR_\mathrm{s}$, $T_{\mathrm{i}}^\mathrm{max}$ of the M-thin disc
reaches $3.29\times 10^8$K, $1.31\times 10^9$K, $6.83\times 10^9$K for
$\dot{m} = 10^{-4}, 10^{-3}, 10^{-2}$, respectively, at $R=7.9
R_\mathrm{s}$. As for ADAFs, $T_{\mathrm{i}}^\mathrm{max}$ remains
$8.8\times 10^{10}$K at $R=5.4 R_\mathrm{s}$, regardless of
the variation of the accretion rate.

\begin{figure} 
	\includegraphics[width=\columnwidth]{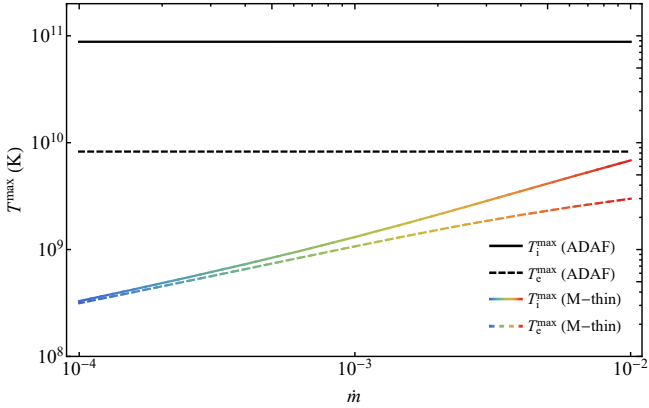}
  \caption{The maximum ion temperature $T_{\mathrm{i}}^\mathrm{max}$
  (solid) and the maximum electron temperature
  $T_{\mathrm{e}}^\mathrm{max}$ (dashed) as a function of the
  accretion rate $\dot{m}$ for ADAFs ($M=10M_\odot$, black) and M-thin
  discs ($M=1.4M_\odot$, rainbow changing from blue to red as the
  temperature increases), where $\alpha=0.1$, $j=1.8cR_\mathrm{s}$ in
  both models.}
  \label{Fig:T-m}
\end{figure}

Specifically in our calculation, $T_{\mathrm{i}}^\mathrm{max}$ and
$T_{\mathrm{e}}^\mathrm{max}$ of ADAFs are nearly independent of
$\dot{m}$, whereas the relation between $T_{\mathrm{i}}^\mathrm{max}$
and $\dot{m}$ in M-thin discs is close to
$T_{\mathrm{i}}^\mathrm{max}\propto \dot{m}^{0.7}$, which is presented
in Figure~\ref{Fig:T-m}. Therefore, for clarity we only show the
temperature distribution of ADAFs calculated by $\dot{m}_{*} =
10^{-4}$ in Figure~\ref{Fig:T-R}. Nevertheless, we point out that the
value of $\alpha$ and $j$ can significantly change the temperature
properties. For example, if we take $j=1.5cR_\mathrm{s}$ and
$\alpha=0.3$, $T_{\mathrm{i}}^\mathrm{max}$ of M-thin discs will reach
$1.74\times 10^8$K, $6.19\times 10^8$K, $2.87\times 10^9$K for
$\dot{m}= 10^{-4}$, 10$^{-3}$, 10$^{-2}$, respectively, at $R=3
R_\mathrm{s}$, and $T_{\mathrm{i}}^\mathrm{max}$ of ADAFs will remain
$3.77\times 10^{11}$K at $R=3 R_\mathrm{s}$.

\subsection{Theoretical regions}

Now we can solve out the X-ray and radio luminosity theoretically when
$M, \alpha, j, \dot{m}$ are given. Considering the observational data
are scattered in some regions, it is reasonable for us to change above
parameters within suitable ranges to create theoretical counterparts,
comparing them with the observational data to verify our models.

\begin{figure*}
	\includegraphics[width=\textwidth]{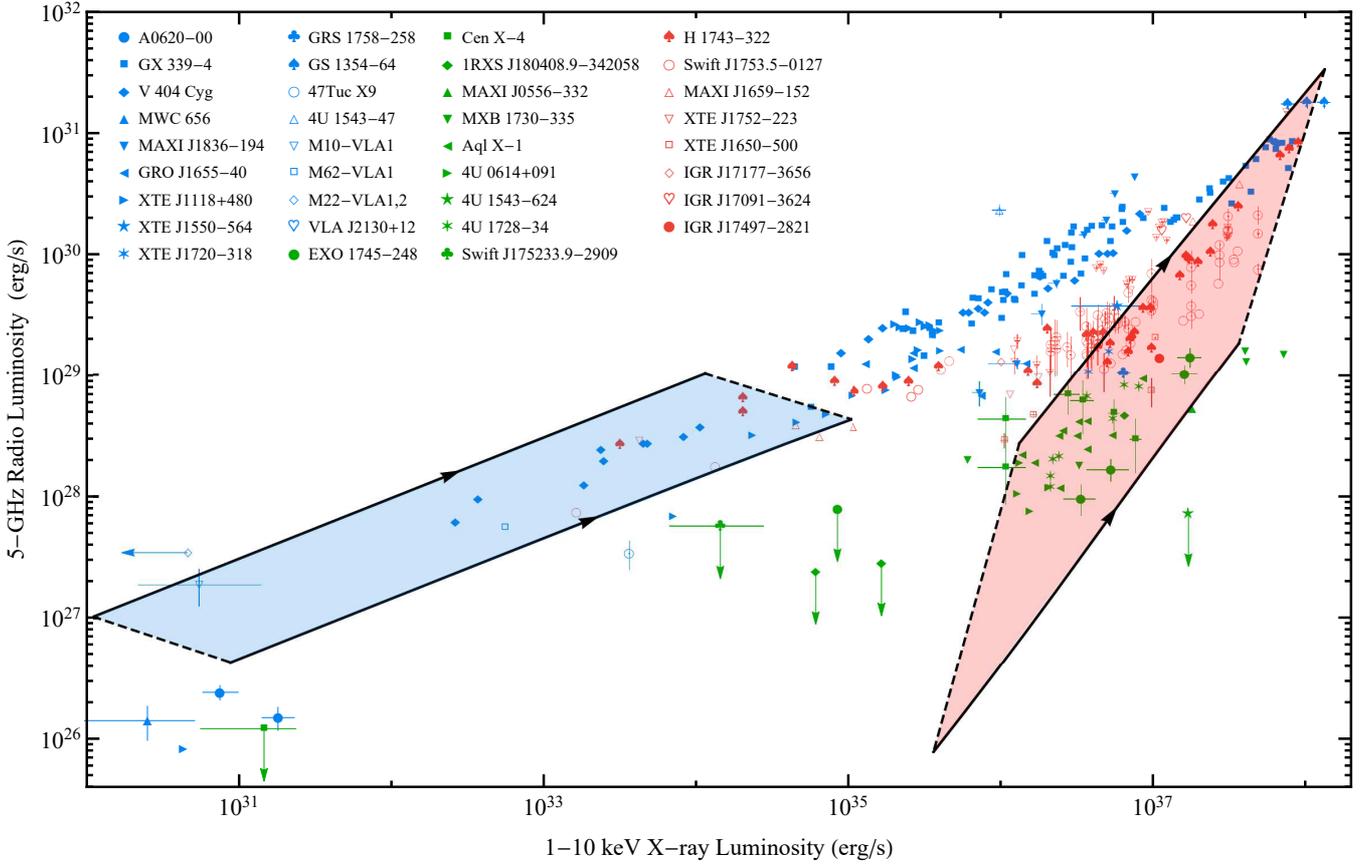}
  \caption{Radio (5 GHz) and X-ray (1-10 keV) luminosities for BHXRBs
  and NSXRBs (see Table~\ref{Tab:Data} for references). Blue dots are
  data of BHXRBs following the standard radio/X-ray correlation; Red
  dots are data of BHXRB outliers; Green dots are data of
  non-pulsating NSXRBs. The blue region is calculated by ADAFs with
  $M=10M_\odot$, $\dot{m}_*= 10^{-4} - 10^{-2}$, where $\alpha= 0.3$,
  $j=1.5cR_\mathrm{s}$ are given in the upper boundary and $\alpha=
  0.1$, $j=1.8cR_\mathrm{s}$ are given in the lower boundary, showing
  a slope of $\sim0.5$; The red region is calculated by the M-thin
  disc with $\dot{m}= 10^{-3} - 10^{-1}$, where $M=10M_\odot$,
  $\alpha= 0.01$, $j=1.8cR_\mathrm{s}$ are given in the upper boundary
  for BHXRBs outliers and $M=1.4M_\odot$, $\alpha= 0.3$,
  $j=1.5cR_\mathrm{s}$ are given in the lower boundary for NSXRBs,
  showing a slope of $\sim 1.6$. Black arrows indicate the increase of
  the accretion rate; Colorful arrows denote the upper limits of the
  observational data.}
  \label{Fig:R-X}
\end{figure*}

Figure~\ref{Fig:R-X} illustrates two theoretical regions
obtained by ADAFs and M-thin discs with different parametric ranges,
whose black arrows in the upper/lower boundaries indicate the increase
of the accretion rate. The blue region is calculated by ADAFs with
$M=10M_\odot$ to explain the standard BHXRBs radio/X-ray correlation.
In order to create a theoretical region, we firstly take $\alpha =
0.3$ and $j=1.5cR_\mathrm{s}$ into the calculation, together with
$\dot{m}_*$ changing from $10^{-4}$ to $10^{-2}$. The results are
shown as the upper boundary of the blue region, where the radio and
X-ray luminosities both increase as $\dot{m}_*$ goes up. Then we
change $\alpha$ and $j$ into $0.1$ and $1.8cR_\mathrm{s}$, the
corresponding results are given in the lower boundary of the region.
Through adjusting $\alpha$ and $j$ linearly, we link the
aforementioned two upper/lower boundaries with two dashed lines, which
forms the theoretical ADAF region. Analogously, we can obtain the red
region for BHXRBs outliers and NSXRBs by M-thin discs with $\dot{m}=
10^{-3} - 10^{-1}$, where $M=10M_\odot$, $\alpha= 0.01$,
$j=1.8cR_\mathrm{s}$ are given in the upper boundary for BHXRBs
outliers, and $M=1.4M_\odot$, $\alpha= 0.3$, $j=1.5cR_\mathrm{s}$ are
set in the lower boundary for NSXRBs. Not only do the theoretical
results of ADAFs and M-thin discs lie in different parts of the plane,
but also they exhibit different slopes apparently. Our numerical
solutions give 0.49 for ADAFs and 1.68 for M-thin discs, which are
consistent with previous observational data fitting results within the
range of errors.

\subsection{Observations and correlations}

Since the theoretical regions have already been obtained, we
are able to introduce the simultaneous radio and X-ray observational
data to verify our model assumptions. We adopt 35 XRB sources from the
online repository \citep{ArashBahramian18}, where all the measurements
are converted to 5 GHz (in radio) and 1-10 keV (in X-rays),
respectively, by assuming a flat radio spectral index and a power-law
model in X-rays. For the standard radio/X-ray correlation of BHXRBs,
we choose 17 sources: A0620-00, GX339-4, V404Cyg, MWC 656, MAXI
J1836-194, GRO J1655-40, XTE J1118+480, XTE J1550-564, XTE J1720-318,
GRS 1758-258, GS 1354-64, 47Tuc X9, 4U 1543-47, M10-VLA1, M62-VLA1,
M22-VLA1,2 and VLA J2130+12, whose data are ranging from LHS to QS. As
for NSXRBs, we only consider 10 non-pulsating sources: EXO 1745-248,
Cen X-4, 1RXS J180408.9-342058, MAXI J0556-332, MXB 1730-335, Aql X-1,
4U 0614+091, 4U 1543-624, 4U 1728-34 and Swift J175233.9-2909. In
terms of BHXRB outliers, we collect 8 sources, including H1743-322,
Swift J1753.5-0127, MAXI J1659-152, IGR J17091-3624, IGR J17177-3656,
IGR J17497-2821, XTE J1650-500 and XTE J1752-223. Their corresponding
references are listed individually in Table~\ref{Tab:Data}.

Figure~\ref{Fig:R-X} denotes three types of sources in three colours:
blue dots are data of BHXRBs following the standard radio/X-ray
correlation; red dots are data of BHXRB outliers; green dots are data
of non-pulsating NSXRBs. Different types of data lie well
inside the corresponding theoretical regions and show nearly the same
slope as the theoretical expectation. However, due to the fact that
there is no solution for ADAFs for large radii if $\dot{m}>0.01$, our
blue region cannot cover data of standard BHXRBs at high luminosity,
which indicates that the ADAF is applicable to the QS and the LHS. As
for BHXRBs at high X-ray luminosity, the LHAF model may be a promising
model to understand their behaviour \citep{Coriat11, Koljonen19}.

\section{Conclusions and discussion} \label{Sec: Conclusions}

\begin{table*}
  \caption{\label{Tab:Data} References of 35 simultaneous radio and X-ray sources}
  \begin{tabular}{cccc}
    \hline
    Sources & Types & References \\  \hline
    A0620-00 & BHXRB & \citealt{Gallo06, Dincer17, Jonker04} \\
    GX 339-4 & BHXRB & \citealt{Corbel13, Zdziarski04} \\
    V 404 Cyg & BHXRB & \citealt{Corbel08, Miller-Jones09, Rana16, Plotkin17} \\
    MWC 656 & BHXRB & \citealt{Ribo17} \\
    MAXI J1836-194 & BHXRB & \citealt{Russell15} \\
    GRO J1655-40 & BHXRB & \citealt{Coriat10, Calvelo10} \\
    XTE J1118+480 & BHXRB & \citealt{Fender10, Gallo14} \\
    XTE J1550-564 & BHXRB & \citealt{Gallo03} \\
    XTE J1720-318 & BHXRB & \citealt{Brocksopp05} \\
    GRS 1758-258 & BHXRB & \citealt{Gallo03} \\
    GS 1354-64 & BHXRB & \citealt{Gallo03} \\
    47Tuc X9 & BHXRB & \citealt{Miller-Jones15, Bahramian17} \\
    4U 1543-47 & BHXRB & \citealt{Gallo03} \\
    M10-VLA1 & BHXRB & \citealt{Shishkovsky18}. \\
    M62-VLA1 & BHXRB & \citealt{Chomiuk13} \\
    M22-VLA1,2 & BHXRB & \citealt{Strader12} \\
    VLA J2130+12 & BHXRB & \citealt{Tetarenko16a} \\
    EXO 1745-248 & NSXRB & \citealt{Valenti07, Tetarenko16} \\
    Cen X-4 & NSXRB & \citealt{Tudor17} \\
    1RXS J180408.9-342058 & NSXRB & \citealt{Gusinskaia17} \\
    MAXI J0556-332 & NSXRB & \citealt{Coriat11a} \\
    MXB 1730-335 & NSXRB & \citealt{Rutledge98} \\
    Aql X-1 & NSXRB & \citealt{Jonker04, Tudose09, Miller-Jones10, Migliari11} \\
    4U 0614+091 & NSXRB & \citealt{Migliari11} \\
    4U 1543-624 & NSXRB & \citealt{Ludlam17} \\
    4U 1728-34 & NSXRB & \citealt{Migliari03, Galloway03, Migliari11} \\
    Swift J175233.9-2909 & NSXRB & \citealt{Tetarenko17} \\
    H 1743-322 & BH Outlier & \citealt{Coriat11} \\
    Swift J1753.5-0127 & BH Outlier & \citealt{Zurita08, Soleri10, Rushton16, Plotkin17} \\
    MAXI J1659-152 & BH Outlier & \citealt{Jonker12} \\
    XTE J1752-223 & BH Outlier & \citealt{Ratti12, Brocksopp13} \\
    XTE J1650-500 & BH Outlier & \citealt{Corbel04} \\
    IGR J17177-3656 & BH Outlier & \citealt{Paizis11} \\
    IGR J17091-3624 & BH Outlier & \citealt{Rodriguez11} \\
    IGR J17497-2821 & BH Outlier & \citealt{Rodriguez07} \\
    \hline
  \end{tabular}
  \end{table*}

In this work, we have applied the magnetized, optically and
geometrically thin, two-temperature, radiative cooling-dominated
accretion disc (the M-thin disc) to BHXRB outliers and
non-pulsating NSXRBs. We have also calculated the ADAF for
standard BHXRBs. Firstly, we have derived the ion and electron
temperatures as a function of radius and accretion rate for both
models. Then we investigated the relations between radio luminosity
and temperatures, and assumed that the jet power is offered by a
fraction of the internal energy from the accretion disc. We used the
maximum ion temperature for an estimation of the internal energy rate
and calculated the theoretical X-ray and radio luminosity. Moreover,
we have also deduced the powerlaw indexes 0.5 and 1.5 for BHXRBs and
NSXRBs via ADAFs and M-thin discs, respectively, and have
found the two different theoretical regions well match the
observational data. Therefore, the black hole system may have two
choices, i.e., either the ADAF or the M-thin disc, whose slope
corresponds to $\sim 0.5$ (radiatively inefficient) and $\sim 1.5$
(radiatively efficient), respectively.

Recently, \citet{Tremou20} reported new simultaneous
radio/X-ray measurements of the quiescent GX 339-4, which well lie
inside the ADAF region in Figure~\ref{Fig:R-X}. Nonetheless at the
same time, \citet{Yan20} found an unexpected ``cooler when fainter''
(positive $T_\mathrm{e}-L_{\mathrm{X}}$ correlation) branch in the
low-luminosity regime ($L_{\mathrm{X}} < 3\times 10^{36}$
erg~s$^{-1}$), which puts a challenge to the classic ADAF model. On
the other hand, as shown by our Figure~\ref{Fig:T-R}, the M-thin disc
may well explain such a correlation. 

Nevertheless, we should point out that in order to simplify the
analysis and the calculation, some simple assumptions have been made
in our models. For instance, several parameters (e.g., the
magnetic pressure $\beta$, the transferred fraction $f^*$) are assumed
to be constant, together with a flat spectral index in the
observational data. It is worth evaluating our models under more
variables and a non-flat spectral index \citep{Espinasse18} in further
studies; Except for the power of jets, there are other factors (e.g.,
the mass of the compact object in \citealt{Kording06}, radiative
efficiency in \citealt{Yuan14}, inclination of the binary system in
\citealt{Motta18}) could mildly affect the radio luminosity, and it
would be interesting to take these factors into consideration in
future works; Only the bremsstrahlung cooling process is taken into
account, which indicates temperatures in the real cases may be lower
than the current solutions. In addition, we should note that the M-thin
disc model is limited to the non-pulsating NSXRBs, while it is of great
interest to develop new accretion disc models (e.g, the magnetic
flux-dominated jet model proposed by \citealt{Parfrey16}) to
understand the disc-jet couplings in pulsating NSXRBs.

However, in our opinion, these assumptions are valid for us to make a
comparison between theoretical results and observational data over
orders of magnitude, delivering fruitful insights of the disc-jet
couplings of BHXRBs and non-pulsating NSXRBs, along with revealing the
potential connections between the jet power and the internal energy of
the inner accretion disc.

\section*{Acknowledgements}

We thank Zhen Yan, Hui Zhang, Shan-Shan Weng, Yi-Ze Dong, Hao He,
and Bo-Cheng Zhu for beneficial suggestions, and the anonymous referee
for providing very useful comments that helped improve this paper.
This work was supported by the National Natural Science Foundation of China
under grants 11925301 and 11573023.

\bibliographystyle{mnras}

\label{lastpage}
\end{document}